# *b**AI**oimage analysis: elevating the rate of scientific discovery — as a community*

Damian Edward Dalle Nogare[1], Matthew Hartley[2], Joran Deschamps[1], Jan Ellenberg[3], Florian Jug[1]

[1] Fondazione Human Technopole, Milan, Italy
[2] EMBL-EBI, Hinxton, UK
[3] EMBL Heidelberg, Heidelberg, Germany

**ABSTRACT**   The future of bioimage analysis is increasingly defined by the development and use of tools that rely on deep learning and artificial intelligence (AI). For this trend to continue in a way most useful for stimulating scientific progress, it will require our multidisciplinary community to work together, establish FAIR data sharing and deliver usable, reproducible analytical tools.

Bioimage analysis is in the midst of a revolution that will profoundly shape the future of the field for decades, spurred by recent developments in deep learning and artificial intelligence. When thinking about the future of this change, it is tempting to imagine one where current limitations and pain-points have been resolved. Although imagining such a future is easy, it is less easy to imagine the transition from today's status quo to that desired future state — what it will require, and how it will be enabled. In this commentary, we discuss how to facilitate scientific progress in the life-sciences, identify necessary changes, and provide ideas for how these changes might best be realized.

Over the past decade, the use of artificial intelligence has revolutionized bioimage analysis. The number of results in a PubMed search of the biomedical literature for the phrase "deep learning" increased exponentially between 2012 and 2022, and the number of articles indexed by PubMed that mention deep learning more than doubled between 2020 and 2022 alone (from 9,303 in 2020 vs. 19,650 in 2022). State-of-the-art tools for many common bioimage analysis tasks such as segmentation and denoising, which are critical for generating scientific insight from raw image data, now employ AI. For many use-cases these tools significantly outperform their classical competitors in speed and accuracy. Such tools are significant force-multipliers for biological discovery, facilitating advances that would be difficult or impossible with more classical tools that do not rely on AI. There is vast potential waiting to be unlocked in many areas such as computational microscopy, multi-modal data analysis, cell tracking, phenotypic classification, and smart microscopy, and our community must find efficient ways to enable this potential as quickly as possible.

## Challenges for AI in bioimage analysis

Over the next ten years, we anticipate two major challenges for the development of AI in bioimage analysis, from the perspective of users as well as developers of AI tools. As we will see, these two challenges are deeply intertwined, as they both stem from the fact that the performance of AI-based methods and tools is closely tied to the data that was used to train the final model.

From a method developer perspective, as discussed below, a wider range of open and standardized data, metadata, and ground truth labels need to become available in order to advance the state of the art. These data should be chosen or generated such that it enables method developers to tackle challenging analyses that are current impediments to scientific progress in the life sciences.

From a user perspective, finding appropriate models to analyze a dataset is currently not a straight-forward task. Even if many models are publicly available to users, choosing a suitable one requires a way to evaluate the quality of model predictions on their data. Indeed, predictions generated by a given model need to be critically assessed and carefully interpreted to ensure the responsible use of AI-driven tools. This can be enabled by offering suitable training opportunities and providing tools that deliver interpretable quality metrics.

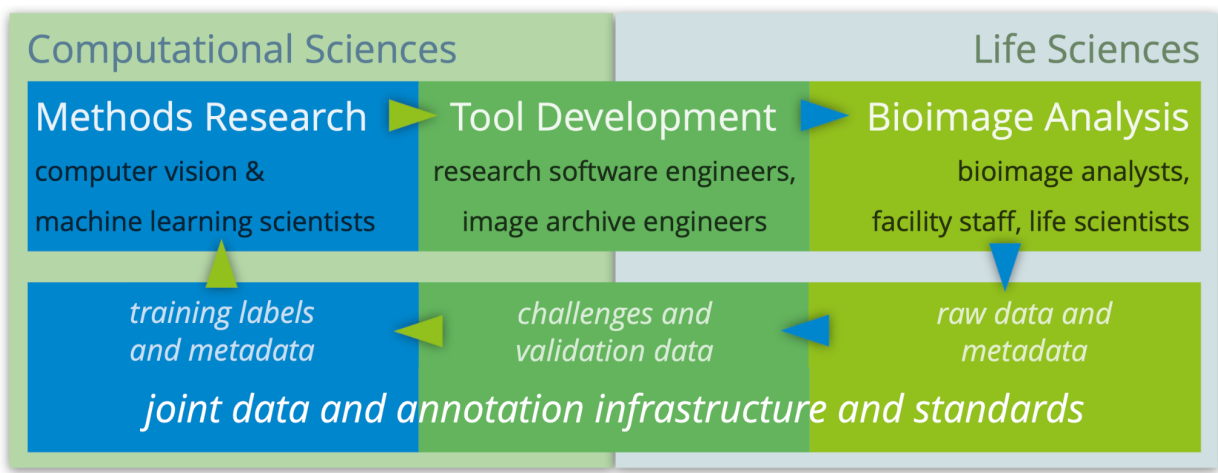

*Figure 1:* The field of bioimage analysis will thrive if the disparate parts of our community, each with their own responsibilities, work together toward the common goal of elevating the rate of scientific discovery in the life sciences. This could be facilitated by a suitable joint data and annotation infrastructure and standards on which the community agrees.

## Ground-breaking AI research requires a fair amount of FAIR data

The reason that AI-based methods outperform classical approaches in so many analysis tasks is that they can distill the most relevant priors[1] from a given body of training data. Trained networks are therefore precisely tailored to solving a specific analysis task in the context of a specific type of data, *i.e.,* the kind of data on which they were trained. These training data must be of sufficient quality and quantity, and more importantly, paired with high-quality expert annotations. In addition to limitations around training data, there is currently an unmet need for reference datasets that can be used to compare and benchmark the performance of the rapidly growing number of tools for common bioimage analysis applications. Some benchmark datasets exist today but despite their undisputed utility, they are of vastly diverging quality, age, and practical relevance, and do not share a common standard for storage and accessibility. This lack of common standards makes it difficult to evaluate computational tools across multiple reference datasets, rendering it much harder to develop widely generalizable techniques. Better reference datasets and tool comparisons will enable life scientists to determine which of the rapidly growing zoo of methods can best solve a given bioimage analysis problem.

---

[1] A data prior is a task-specific clue extracted from previously seen examples that can later be used to make better decisions when new data is processed.

These demands for large amounts of well-annotated and structured data require consensus on how data collection, annotation, storage, and access can best be unified and facilitated. Moving forward, we will need to make decisions as a community about how to address these challenges. Unfortunately, there is a natural limit to the extent to which individual AI researchers can achieve this goal. While major breakthroughs in the next 10 years will certainly require advancements in the methodology and computational frameworks that drive modern AI tools, they will equally require better ways to generate, use, and share data. The latter necessitates solving the technical challenges of storing and effectively sharing large datasets, but also reaching community consensus on agreeable formats for images, image metadata, and annotations like ontologies and ground-truth label data. Needless to say, this also relies on finding or establishing suitable, stable, and long-term funding sources to develop and maintain the required infrastructure.

We also need to support, encourage, and incentivize widespread data annotation, sharing, and reuse. The FAIR — Findable, Accessible, Interoperable, Reusable — principles were developed in large part to address these challenges and are a core part of the solution, also for image data. In the context of AI-driven bioimage analysis, publicly available FAIR data allows the community to document the key analysis needs, and enables the creation of better methods, method evaluation, and user-facing tools — ultimately supporting the goal of increasing the rate of scientific discovery.

## Life scientists and method developers — better together!

A strengthened collaborative partnership between life scientists and method developers that addresses the challenges outlined above should lead to a positive feedback loop of accelerated technology development and successful application. However, such a partnership is not without its challenges. Life scientists generate large amounts of raw image data and are in many cases the only ones capable of providing expert annotations. Thus, they are key partners in advancing the field of bioimage analysis. Unfortunately, the effort of annotating and depositing new image data in a FAIR-compliant way is, if currently possible at all[1,2], significant and in many cases, largely unrewarded. Hence, we should improve data submission procedures for existing (or newly created) image archiving infrastructures such that data sharing becomes as technically frictionless as possible. Just as importantly, AI researchers need to invest the time and effort required to transform their methods into easily usable tools that address the analysis needs of life scientists[3]. Unfortunately, such efforts also often go professionally unrewarded. Although there might be indirect rewards for life scientists and method developers who operate this way, the scientific community must also strive to create additional incentives and reward structures. The easiest and most immediate action would be to strengthen publication and citation of computational tools and establish a concept of data-citations that would integrate with existing scientific success metrics and could be used by hiring, promotion, and tenure committees as well as grant review panels. While technically easy to implement, broad acceptance of such citations as representative of scientific output is a challenging social problem without an easy solution[4]. More effective translation from methods to tools can also be achieved if our communities hire and support more research software engineers and bioimage analysts in a professional capacity, since they can incorporate the latest methods into usable software tools and best mediate directly with bench scientists to apply new analysis tools successfully. Only close collaboration and partnership between computational and life scientists will foster creation of a forward-looking system that more rapidly and efficiently facilitates scientific discoveries.

This facilitation of scientific discovery, though, requires realistic expectations among users about what AI models can, and cannot, predict, in order for bioimage analysis to remain rigorous and reproducible. For example, AI cannot precisely recover fine details of structures in diffraction limited microscopy images that are below the diffraction limit of the microscope, as information at those spatial scales is lost during image formation[5]. At best, AI can make predictions about what such structures might look like, based on the raw input image and a data prior that was previously distilled from the available body of training data. Segmentation methods, for instance, learn to do a good job on challenging parts of the image by incorporating a learned prior on the typical shapes or textures of objects to be segmented. Furthermore, the outputs of given models themselves deal with uncertainty in different ways, and no universally applicable quality metric does exist[6]. In some cases (e.g., in recent denoising approaches[7]), what is returned from an AI model are sampled "interpretations" of the given raw input, drawn from a previously learned distribution of reasonable data appearances. Many approaches, however, return a single output[8], which most often is something closer to the "average" of all possible denoised interpretations.

While these issues are known to core method developers, they might not be as well understood by users, and how best to deal with them is not trivial and remains a topic of active discussion in the AI community. This underscores the importance of an open discourse and consistent training efforts in this area and, in the longer run, of broadly accepted standards and quality metrics for the predictions of future analysis tools. The goal must be to enable the life science community to identify the best method or tool for a given job and to discriminate not only qualitatively but also quantitatively to what extent predictions can be trusted to infer facts about the underlying biology. This will eventually help to alleviate the use of impenetrable AI black-boxes within scientific data analysis pipelines.

Given the uncertainties regarding the quality of predictions and the dependence on suitable training data, a key challenge is validation and reproducibility of reported results. This is an innocent looking but difficult problem that is not easy to solve in full generality, but one that will also benefit tremendously from FAIR data resources, open sharing, standardized test datasets, and a simplified way to compare analysis tools.

## The goal-oriented collaborative future of bioimage analysis is bright

A collaborative partnership between life scientists and method developers is a two-way street between the biological questions being asked and dedicated and targeted methodologies being created. This requires bioimage analysts, data stewards, data scientists, and research software and data archive engineers, acting as the glue between users and developers in our community, coming together to build an open and FAIR data infrastructure (see Figure 1). Data and data annotations should be high quality and represent the types of problems that currently limit the rate of scientific progress in order to enable method developers to work on the most impactful problems. At the same time, models need to be shared openly and be sufficiently documented, reusable, and quantitatively assessable. This approach will be key to synergistically elevate the rate of scientific progress in the life sciences and in AI-based bioimage analysis research.